\documentclass[12pt,nofootinbib]{revtex4}

\newcommand{\sg}{\sigma}
\newcommand{\tw}{\tilde{\omega}}
\newcommand{\bz}{\bar{z}}
\newcommand{\half}{\frac{1}{2}}
\newcommand{\bea}{\begin{eqnarray}}
\newcommand{\eea}{\end{eqnarray}}
\newcommand{\beq}{\begin{equation}}

\newcommand{\eeq}{\end{equation}}

\begin{document}

\title{Noncommutative Dirac oscillator in  an external magnetic field}


\author{Bhabani Prasad Mandal \footnote{e-mail address:
\ \ bhabani@bhu.ac.in, \ \ bhabani.mandal@gmail.com  } }
\author{Sumit Kumar Rai\footnote{e-mail address: sumitssc@gmail.com} }


\affiliation{ Department of Physics,\\
Banaras Hindu University,\\
Varanasi-221005, INDIA. \\
}
\begin{abstract}
We show that (2+1) dimensional noncommutative Dirac oscillator in an external magnetic field is mapped onto the same but with reduced angular frequency in absence of magnetic field. We construct the relativistic Landau levels by solving corresponding Dirac equation in (2+1) dimensional noncommutative phase space. We observe that lowest Landau levels are exactly same as in commutative space and independent of non-commutative parameter. All the Landau levels become independent of noncommutative parameter for a critical value of the magnetic field. Several other interesting features along with the relevance of such models in the study of atomic transitions in a radiation field have been discussed.
\end{abstract}
\maketitle
The noncommutativity has become a vital field of research owing its development in string theories, quantum field theories and in quantum mechanics \cite{sewi}. The open strings end points  are noncommutative in the presence of the background NS-NS B-field which indicates that the co-ordinates of D-branes are noncommutative \cite{chho1}. There has been a lot of research papers based on perturbative and non-perturbative field theories in noncommutative space \cite{hike}.
An extensive research has also been done on noncommutative quantum mechanical (NCQM) systems \cite{chda,duja,mopo}. In this paper, we study the Dirac oscillator with a perpendicular magnetic field in a noncommutative plane. Noncommutative plane  is characterized by the following commutation relations \cite {napo}
\begin{equation}
\left[x_i,x_j\right]\;=\;i\theta\epsilon_{ij};\quad \quad \left[x_i,p_j\right]=i\hbar\delta_{ij};\quad\quad\left[p_i,p_j\right]=0,\label{ncr}
\end{equation}
where $\theta$ is an arbitrary constant parameter.

Dirac oscillator \cite{mosz} was introduced for the first time by Moshinsky and Szczepaniak (although the similar system was studied much earlier by Ito {\it et al} \cite{itmo}) by adding a term $-imc\omega\beta{{\mbox{\boldmath $\alpha $}}}{\bf \cdot r}$ to the usual Dirac Hamiltonian for a free particle and described by the Hamiltonian
\begin{equation}
H=c {\mbox{\boldmath $\alpha $}}\cdot\left({\bf p}-im\omega\beta{\bf r}\right)+\beta m c^2,
\end{equation}
where ${\bf \alpha}, \beta$ are usual Dirac matrices, m is the rest mass of the particle and c is the speed of light. Initially it was introduced in the context of many body theory, mainly in connection with quark confinement models in quantum chromodynamics (QCD) and in the context of relativistic quantum mechanics \cite{itmo}. In the nonrelativistic limit, the Dirac oscillator reduces to a simple harmonic oscillator with strong spin-orbit coupling term. Because of all these interesting realizations, Dirac oscillator has attracted a lot of attention and has found many physical applications in various branches of physics \cite{lima,noc}.  Most exciting properties of Dirac oscillator lies in its connection with quantum optics \cite{noc} where in 2+1 dimensions it maps to Anti-Jaynes-Cummings (AJC) model \cite{jc} which describes the atomic transitions in a two level system. 

In this present article we have studied Dirac oscillator in noncommutative phase space in the presence of
 an  external magnetic field. The dynamics of the Dirac oscillator in the presence of external uniform magnetic
 field  ${\bf B}$ in noncommutative phase space is  governed by the Hamiltonian,
\begin{equation}
H= c {\mbox{\boldmath $\alpha $}}\cdot ({\bf p}-\frac{e{\bf A}}{c} -im\omega\beta {\bf r}) + \beta m c^2
\label{dom}.
\end{equation}
${\bf A}$ is the vector potential which produces the external magnetic field
and $e$ is the charge of the Dirac oscillator which is considered to be an
 electron in this case.  This theory of (2+1) dimensional Dirac oscillator in noncommutative phase space with an external magnetic field is mapped onto the theory of 2+1 dimensional Dirac oscillator without magnetic field in the same noncommutative phase space given by the Hamiltonian
 
\begin{equation}
H^\prime= c{\mbox{\boldmath$\alpha$}}\cdot ({\bf p} -im\tw\beta {\bf r}) + \beta m c^2, {\mbox where}\;\; \tw =\omega - \frac{\omega_c}{2}.
\label{dowm}
\end{equation}
The constant magnetic field decreases the angular frequency of the Dirac oscillator by half of the cyclotron frequency ($ \omega_c=\frac{B\left|e\right|}{mc}$). For the sake of simplicity, we consider the magnetic field, ${\bf B}$
along the $z$-direction.
The  vector potential for this particular magnetic field
can be chosen in the  symmetric gauge as
 ${\bf A}= ( -\frac{B}{2}y,
\frac{B}{2}x, 0).$ We solve the Dirac equation corresponding to this Hamiltonian in noncommutative phase space to obtain the energy eigenvalues and eigenfunctions analytically. The lowest Landau levels (LLL) are constructed explicitly in non-commutative space which is exactly same as the LLL obtained in commutative space \cite{mave}. Further, we also show the exact mapping of this relativistic model to the AJC model which is so widely used in quantum optics. This model shows properties similar to the properties of graphene \cite{grar}, monolayer of carbon atoms.

We are  studying the system in two space
and one time dimension
 where  Dirac matrices can be chosen  in terms of
$2\times 2 $ Pauli spin  matrices. We consider one of the widely used choices 
of Dirac matrices as
    $\alpha_x =\sg_x , \alpha_y = \sg_y \ $ and $ \ \beta =\sg_z .$
 The  Dirac Hamiltonian in  noncommutative space given by Eq. (\ref{dowm}) can be written nicely in terms of complex
 coordinate
$z=x+iy$ and conjugate momentum in noncommutative phase space $\Pi_z$, as
\begin{equation}
H=\left ( \begin{array}{c c} mc^2 & 2c\Pi_z+ic\left(m\tw +\frac{\hbar}{\theta}\right)\bz \\2c\Pi_{\bz}- ic\left(m\tw +\frac{\hbar}{\theta}\right)z & -mc^2 \end{array} \right )
\label{ham},
\end{equation}
where the conjugate momentum in noncommutative phase space is defined as  \cite{napo}
\begin{eqnarray}
\Pi_z&=&p_z-\frac{i\hbar}{2\theta}{\bar{z}}\nonumber\\
\Pi_{\bar{z}}&=&p_{\bar{z}}+\frac{i\hbar}{2\theta}z,
\end{eqnarray}
where
\begin{eqnarray}
 p_z &=&-i\hbar\frac{d}{dz} = \frac{1}{2}(p_x-ip_y) \nonumber ,\\
 p_{\bz}&=&- i\hbar\frac{d}{d\bz} = \frac{1}{2}(p_x+ip_y),
\label{pz1} \\
\end{eqnarray}
which satisfy $\left[z,p_z\right]=i\hbar=\left[\bar{z},p_{\bar{z}}\right]$ and $\left[z,p_{\bar{z}}\right]=0=\left[\bar{z},p_{z}\right]$.
 The noncommutative  algebra in Eq. (\ref{ncr}) can then be written as
 \begin{eqnarray}
&&\left[z,\bar{z}\right]\;=\;2\theta;\quad \quad \left[z,\Pi_{\bar{z}}\right]=0=\left[\bar{z},\Pi_z\right];\quad\quad\left[\Pi_z,\Pi_{\bar{z}}\right]=\frac{\hbar^2}{2\theta},\nonumber\\
&&\left[z,\Pi_z\right]\;=\;\left[\bar{z},\Pi_{\bar{z}}\right]\;=\;0.
\end{eqnarray}

The  time indepenent planar  Dirac equation ($H\left|\psi\right\rangle=E\left|\psi\right\rangle$) corresponding to  the Hamiltonian given in Eq. (\ref{ham}), can then be written in component form as \begin{eqnarray}
\left(E-mc^2\right)\left|\psi_1\right\rangle &=&c\left[2\Pi_z+i\left(m\tw +\frac{\hbar}{\theta}\right)\bz\right]\left|\psi_2\right\rangle,\nonumber\\
\left(E+mc^2\right)\left|\psi_2\right\rangle &=&c\left[2\Pi_{\bz}- i\left(m\tw +\frac{\hbar}{\theta}\right)z\right]\left|\psi_1\right\rangle,\label{ce}
\end{eqnarray}
where two component $\left|\psi\right\rangle=\left ( \begin{array}{c c} \left|\psi_1\right\rangle 
\\ \left|\psi_2\right\rangle  \end{array} \right )$.
Now we introduce the following two sets of   operators
\begin{eqnarray}
a&=&\frac{1}{\sqrt{2\theta}}\;z\;;\quad\quad a^\dagger=\frac{1}{\sqrt{2\theta}}\;\bar{z}\;;\nonumber\\
b&=&\sqrt{\frac{2\theta}{\hbar^2}}\;\Pi_{z}\;;\quad\quad b^\dagger=\sqrt{\frac{2\theta}{\hbar^2}}\;\Pi_{\bar{z}}\;.
\end{eqnarray}
It can be checked easily that $a,\;a^\dagger,\;b $ and $ b^\dagger$ satisfy the following algebra
\begin{equation}
\left[a,a^\dagger\right]=1,\quad\quad \left[b,b^\dagger\right]=1,
\end{equation}
where as other commutators are zero.
Using these creation and annihilation operators, the coupled equation given by Eq. (\ref{ce}) can be expressed as 
\begin{eqnarray}
 \left(E-mc^2\right)\left|\psi_1\right\rangle&=&ig\left(a^\dagger \cosh\Phi-ib \sinh\Phi\right)\left|\psi_2\right\rangle,\nonumber\\
 \left(E+mc^2\right)\left|\psi_2\right\rangle&=&-ig\left(a \cosh\Phi+ib^\dagger \sinh\Phi\right)\left|\psi_1\right\rangle,\label{ce2}
\end{eqnarray}
where 
\begin{equation}
g=c\sqrt{\frac{2}{\theta}\left[{\left(m\tilde\omega\theta+\hbar\right)}^2-\hbar^2\right]}\;\; {\mbox{and}}\;\; \tanh\Phi=\frac{\hbar}{m\tilde\omega\theta+\hbar}.\label{geq}
\end{equation}
It is interesting to note that for a value of noncommutative parameter $\theta=\frac{2\hbar}{m\tilde\omega}$, the coupling constant $g=\sqrt{2}\;g_c$, where $g_c$ is the coupling constant in the commutative case \cite{mave}.
In order to find energy spectra, we further define annihilation and creation operators as follows
\begin{eqnarray}
C&=&a \cosh\Phi+ib^\dagger \sinh\Phi\nonumber\\
C^\dagger&=&a^\dagger \cosh\Phi-ib \sinh\Phi,\label{anc}
\end{eqnarray}
satisfying  the commutation relation as
\begin{equation}
\left[C,C^\dagger\right]=1.
\end{equation}
The  Eq. (\ref{ce2}) can be decoupled  as  
\begin{eqnarray}
\left(E^2-m^2c^4\right)\left|\psi_1\right\rangle &=&g^2C^\dagger C\left|\psi_1\right\rangle,\nonumber\\
\left(E^2-m^2c^4\right)\left|\psi_2\right\rangle &=&g^2\left(C^\dagger C+1\right)\left|\psi_2\right\rangle,\label{dce}
\end{eqnarray}
Writing the spinors wave function in basis of number operators $n=C^\dagger C$ as
\begin{eqnarray}
\left(E^2-m^2c^4\right)\left|n\right\rangle &=&g^2n\left|n\right\rangle,\nonumber\\
\left(E^2-m^2c^4\right)\left|n^\prime\right\rangle &=&g^2\left(n^\prime+1\right)\left|n^\prime\right\rangle,\label{eee}
\end{eqnarray}
The relativistic Landau levels or the energy eigenvalues so obtained from Eq. (\ref{eee}) are as follows
\begin{equation}
E=\pm E_n=\pm mc^2\sqrt{1+\frac{g^2}{m^2c^4}n}, \quad {\mbox n= 0,1,2,....}\label{eeigen}
\end{equation}
with normalized negative and positive energy states\footnote{We adopt the notation for the state as $|n, \half
m_s>\equiv  \psi_n(z,\bz)\xi_{m_s}$ where $n$ is the
 eigenvalue 
 for the number operator, $C^\dagger
C$ and  $m_s = \pm 1 $ are the eigenvalues of the operator $\sg_z$
i.e. $\sg_z\left|\half m_s\right\rangle = m_s\left|\half m_s\right\rangle$, and $\psi_n(z,\bz)$  is the space
part of the wave function in the coordinate representation whereas 
$\xi_{m_s}$ is the spin part of the wave function. These states indicate the
entanglement between orbital and spin degrees of freedom in the Dirac 
oscillator problem.} given by
\beq
\left|\Psi_n^{\pm}\right\rangle\  =\  c_n^{\pm} \left|n; \ \half\right\rangle + d_n^{\pm}\left|n-1;\  -\half\right\rangle,
\label{wf},
\eeq
where
\begin{equation}
c^{\pm}_n =\pm\sqrt{\frac{E_n^+\pm mc^2}{2E_n^+}};\quad {\mbox and}\quad d_n^\pm =\mp\sqrt{\frac{E_n^+\mp mc^2}{2E_n^+}}.\label{cd} 
\end{equation}
From Eq. (\ref{eeigen}), it is clear that energy levels are non-equidistant in energy. For a large value of the magnetic field, $ g$ varies as $\sqrt{|\tilde\omega|}$ and $|\tilde\omega|\propto B$. Therefore, the level separation varies as $\sqrt{B}$.

The lowest Landau levels [LLL] are obtained by using the condition,
\begin{equation}
 C\psi_0(z,\bz) =0.
\end{equation}
 Equivalently  in coordinate representation, the space part  wave function ($\psi_0(z,\bz)$) of the LLL
 satisfies
\begin{equation}
\left(\frac{\partial}{\partial\bz} +\frac{m\tw}{2\hbar}z \right) \psi_0(z,\bz) =0.
\label{lll}
\end{equation}
By substituting
\begin{equation}
 \psi_0(z,\bz) = e^{-\frac{m\tw}{2\hbar} z\bz} u_0(z,\bz).
\label{sub},
\end{equation}
We further obtain
 \begin{equation} 
 \frac{\partial}{\partial\bz}u_0(z,\bz)=0,
\label{lll1}
\end{equation}  as the defining rule for the LLL.
We obtain the LLL in the coordinate representation, which is infinitely degenerate, as
\begin{equation}
\psi_0 (z,\bar{z})=
 z^l e^{-\frac{m\tw}{2\hbar}z\bz},\ \ \ \ l=0,1,2\cdots ,\label{dege1}
\label{lll2}
\end{equation}
 as
$u_0(z,\bz)$ is analytic function and the monomials $z^l $ with $ l=0,1,2 \cdots $
 can  serve as a linearly independent basis. The first excited state and
other higher states in the coordinate space  can be obtained by applying
$C^\dagger $ on the LLL  repeatedly. It is interesting to note that the LLL is independent of noncommutative parameter. 
 A very interesting situation occurs for  a critical value of magnetic field $B=\frac{2\omega mc}{|e|}$ , i.e.  $\tilde{\omega}=0$. The Dirac oscillator stops oscillating and all the Landau levels become independent of $\theta$

Now we would like to show the connection of this model to the AJC model \cite{jc}
so widely used in the study of quantum optics. The Hamiltonian corresponding to
the simple version of AJC model is given as,
\begin{equation}
H_{AJC}= \tilde{g}\sg ^- a +\tilde{g}^*\sg^+a^\dagger + \sg_z mc^2 ,
\label{ajc}
\end{equation}
On the other hand the JC model is described by the Hamiltonian,
\begin{equation}
H_{JC}= \tilde{g}\sg^- a^\dagger + \tilde{g}^*\sg ^+a + \sg_z mc^2 ,
\label{jc}
\end{equation}
which is quite different from Hamiltonian of AJC model as the positions of
creation and annihilation operators are interchanged and defined as $\sigma^{\pm}=\frac{1}{2} \left(\sigma_x\pm i\sigma_y\right) $. $\sigma^+$ and $\sigma^-$ are usual spin raising and lowering
 operators respectively.
The Hamiltonian given in Eq. (\ref{ham}) can be expressed as follows
\begin{equation}
H=\left ( \begin{array}{c c} mc^2 & \tilde{g}C^\dagger \\ \tilde{g}^*C & -mc^2 \end{array} \right )
\label{hamajc},
\end{equation}
where $\tilde{g}=ig$. This Hamiltonian in the Eq. (\ref{hamajc}) can be expressed as 
\begin{equation}
H=\tilde{g}\sigma^+C^\dagger+\tilde{g}^*\sigma^- C+\sigma_z mc^2,
\end{equation}
which is exactly same as $H_{AJC}$  in Eq. (\ref{ajc}). $\tilde{g}$  is the coupling between spin and orbital degrees of freedom in noncommutative phase space which  depends on the strength of the magnetic field as well as on the non-commutative parameter $\theta$.
Hence, it is quite interesting to note that how two different theories are interlinked. This interconnection has been shown for an arbitrary strength of magnetic field and for arbitrary  parameter $\theta$. It provides an alternative approach to study the atomic transitions in two level systems using Dirac oscillator in noncommutative phase space and  in the presence of magnetic field. The connection between quantum optics and relativistics quantum mechanics have been realized experimentally in  \cite {new9} but is yet to be realized in noncommutative phase space.

It is  further shown that the Zitterbewegung frequency for  (2+1)-dimensional Dirac
oscillator in an external magnetic field depends on the strength of the magnetic
field as well as noncommutative parameter $\theta$ when calculated in noncommutative phase space. To show this, we eliminate the state $\left|n-1,-\half\right\rangle$ from  Eq. (\ref{wf}) and then  express the initial pure state at $t=0 $ as
\beq
  \left|n,\half\right\rangle(0)= \frac{E_n^+}{mc^2}\left[d_n^+\left|\Psi_n^-\right\rangle-d_n^-\left|\Psi_n^+\right\rangle\right]
\label{ini}.
\eeq
The Eq. (\ref{ini}) shows that the  starting initial state is 
 a superposition of both  the positive
and negative energy
solutions. The time evolution of this state can be written as
\beq
  \left|n,\half\right\rangle(t)= \frac{E_n^+}{mc^2}\left[d_n^+\left|\Psi_n^-\right\rangle e^{i\omega_nt}-d_n^-\left|\Psi_n^+\right\rangle e^{-i\omega_n t}\right]
\label{tt},
\eeq
where $\omega_n $ is the frequency of oscillation between positive and negative energy
solutions and given as
\beq
\omega_n = \frac{E_n^+}{\hbar} = \frac{mc^2}{\hbar}\sqrt{1+\frac{g^2}{m^2c^4}n}. \label{ww}
\eeq
$\omega_n $ depends on both magnetic field and the non-commutative parameter $\theta$.  
Substituting $\left|\Psi^+\right\rangle$ and $\left|\Psi^-\right\rangle $ and the constants $d^+_n$ and $d^-_n$ in Eq. (\ref{tt}) we obtain,
\bea
\left|n,\frac{1}{2}\right\rangle(t)&=& \left[\cos{(\omega_n t)}-\frac{imc^2}{E_n^+}\sin{(\omega_nt)}\right]\left|n,\half\right\rangle(0)\nonumber \\
&+& i\sqrt{\frac{{E_n^+}^2-m^2c^4}{m^2c^4}}\sin{(\omega_nt)}\left|n-1,-\half\right\rangle(0)\nonumber \\
\eea
This equation shows the oscillatory behavior between the states $|n,\half >$ and 
$\left|n-1,-\half\right\rangle$  which is exactly
similar to atomic Rabi oscillations \cite{jc,rabi} occurring in the JC/AJC models. Rabi frequency is given by
$\omega_n$ given in Eq. (\ref{ww}).
We further observe that the zitterbewegung frequency for the 2+1 dimensional Dirac
oscillator in the external magnetic field depends on the strength of the magnetic
field.

In conclusion, we have studied Dirac oscillator in 2+1 dimension in presence of transverse external magnetic field in non-commutative space by constructing suitable  creation and annihilation operators in terms of properly chosen canonical pairs of coordinates and its corresponding momenta in a complex non-commutative phase space. We have shown that this theory maps onto Dirac oscllator without magnetic field but with different frequency in the same non-commutative space. The frequency is reduced by half of cyclotron frequency ($\omega_c$). We have further solved Dirac equation in 2+1 dimension in non-commutative space to find explicit solutions. Lowest Landau levels are constructed explicitly in the co-ordinate representations which show the infinite degenracy of such state in a simple manner. Landau levels are non-equidistant in energy and the level separation varies as $\sqrt{B}$ for large magnetic field. This feature is quite similar to those shown by graphene. We observe that the LLL are independent of non-commutative parameter $\theta$. For a critical value of magnetic field $B=\frac{2\omega mc}{\left| e\right|}$, the Dirac oscillator stops oscillations and all the Landau levels become independent of $\theta$. The coupling between orbital and spin degrees of freedom depends on the strength of the magnetic field and the noncommutative parameter. Certain specific value of magnetic field, the result reduces to that of Dirac oscillator in commutative plane.  We have further shown that this model can be mapped  onto AJC model, so widely used in the study of quantum optics. Zitterbewegung frequency has been calculated for noncommutative Dirac oscillator in transverse magnetic field. It depends on the strength of magnetic field and noncommutative parameter $\theta$. For large magnetic field it varies as $\omega_n\propto\sqrt{nB}$. This oscillation between negative and positive energy is similar to the Rabi oscillations in the two-level
systems.

\end{document}